# A cell membrane model that reproduces cortical flow-driven cell migration and collective movement

Katsuhiko Sato[1]

[1]Research Institute for Electronic Science, Hokkaido University Sapporo N20W10, 001-0020, Japan

**\* Correspondence:**
Katsuhiko Sato
katsuhiko_sato@es.hokudai.ac.jp

## Abstract

Many fundamental biological processes are dependent on cellular migration. Although the mechanical mechanisms of single-cell migration are relatively well understood, those underlying migration of multiple cells adhered to each other in a cluster, referred to as cluster migration, are poorly understood. A key reason for this knowledge gap is that many forces— including contraction forces from actomyosin networks, hydrostatic pressure from the cytosol, frictional forces from the substrate, and forces from adjacent cells—contribute to cell cluster movement, making it challenging to model, and ultimately elucidate, the final result of these forces. This paper describes a two-dimensional cell membrane model that represents cells on a substrate with polygons and expresses various mechanical forces on the cell surface, keeping these forces balanced at all times by neglecting cell inertia. The model is discrete but equivalent to a continuous model if appropriate replacement rules for cell surface segments are chosen. When cells are given a polarity, expressed by a direction-dependent surface tension reflecting the location dependence of contraction and adhesion on a cell boundary, the cell surface begins to flow from front to rear as a result of force balance. This flow produces unidirectional cell movement, not only for a single cell but also for multiple cells in a cluster, with migration speeds that coincide with analytical results from a continuous model. Further, if the direction of cell polarity is tilted with respect to the cluster center, surface flow induces cell cluster rotation. The reason why this model moves while keeping force balance on cell surface (*i.e.*, under no net forces from outside) is because of the implicit inflow and outflow of cell surface components through the inside of the cell. An analytical formula connecting cell migration speed and turnover rate of cell surface components is presented.

# 1  Introduction

Cellular migration is a key component of numerous biological processes, including the morphogenesis of multicellular organisms, wound healing, and cancer metastasis [1,2,3]. Consequently, elucidating the molecular and biophysical mechanisms that control cell movement can provide fundamental insight to enhance our understanding of these critical processes. Notably, although the mechanisms controlling single-cell migration are relatively well understood [3], those underlying multiple-cell migration, wherein cells adhere to each other and form a cluster, prior to undergoing unidirectional [4,5,6,7,8,9] and rotational motion [10], referred to as cluster migration, remain unclear. One reason for this is that the phenomenon of cluster migration involves what is known as a many-body problem. That is, within cells, there are many forces related to cell movement, such as contraction forces coming from actomyosin, adhesion forces between cells, hydrostatic pressure from cytosol, and forces from adjacent cells; cell movement is a result of balance between these forces. Thus, whereas it is challenging to determine how cell membranes move under this force balance even for a single cell, it is even more difficult in the case of multiple cells, particularly when trying to understand how multiple cells coordinately move. In such complex scenarios, mechanical modeling approaches may be useful, given that these methods guarantee force balance for every element and hence, generate results that can be relied upon with some conviction.

Cell movements are mainly classified into one of two types [11]. In mesenchymal migration, the cell attaches to a substrate at focal adhesions and extends its body at the leading edge by forming lamellipodia and filopodia [12]. During this process, a retrograde flow of actin filaments is observed beneath the plasma membrane of the cell [7,13]. The second type of cell movement is amoeboid migration; in this case, specific adhesion to the substrate is not necessarily required, but, nonspecific friction between the cell membrane and the surrounding matrix is thought to be sufficient for migration [14,15,16,17]. By contracting the rear part of the cell body via a force generated from the actomyosin meshwork under the plasma membrane, the cell increases hydrostatic pressure within the cytosol and produces blebs at the leading edge to extend its body forward [18]. During ameboid migration, cortical flow from front to rear is observed inside the cell, which is believed to play an important role in this process [19]. The shapes of cells undergoing each mode of migration are also different. That is, cells performing amoeboid migration are relatively rounded, whereas those engaging in mesenchymal migration are relatively elongated [7]. Recent studies, however, have begun to suggest that this concept of cell movement involving two distinct migration modes is too limited and does not allow the rigorous classification of all cell movements [3]. Indeed, some cells exhibit both mesenchymal and amoeboid-like movement modes and plastically switch between these, depending on the environmental conditions [20,21]. Thus, some investigators have initiated studies aimed at identifying shared universal mechanisms underlying all cell migration.

From a mechanical modeling viewpoint, a number of common features present in both the mesenchymal and amoeboid modes of migration can be identified. For example, bleb formation in amoeboid migration and lamellipodia formation in mesenchymal migration are similar, given that, in both cases, the membrane in the front region of the cell tends to expand. Mechanically, this behavior of the cell membrane at the leading edge is expressed by weak surface tension. Strong attachment between membrane and substrate in mesenchymal migration and nonspecific friction in amoeboid migration are also expressed by one parameter of a mechanical model. That is, strong attachment is expressed by a large friction coefficient value between membrane and substrate, whereas weak nonspecific friction is expressed by a small friction coefficient value. Moreover, in both modes, contractions at the trailing end of the cell membrane are expressed by a strong surface tension in that region. These observations highlight the common mechanistic features underlying all forms of cell migration. Further, in both migration modes, cortical flow beneath the plasma membrane is present and is thought to play an important role in cell movement [22,23]. However, as noted above, because there are many forces within cells, even if we focus only on the cell membrane, it is very hard to anticipate how cell membranes move and how cells ultimately move eventually under these myriad forces. Therefore, to better understand the mechanical mechanisms underlying cell migration, mechanical models that appropriately express forces within the cells and keep these forces balanced at each point on the cell membrane are needed.

A number of mechanical models for cell migration that satisfy these conditions have been developed. For example, some groups have proposed excellent three-dimensional (3D) surface models, in which directed surface flow and cell division are reproduced [24,25,26]. However, 3D mechanical models require a long computational time and are not appropriate for dealing with multiple cells simultaneously. In 2D cases, cellular vertex models are frequently used for describing multiple cell dynamics, wherein cells are approximated by polygons, and cell boundaries between adjacent cells are represented by straight segments [27]. This model have succeeded in explaining important phenomena in epithelial sheets [27, 29, 30,31]. However, if we extend this 2D cellular vertex model and try to more precisely describe cell surface dynamics on the substrate, some problems arise. One minor problem is that because the cell boundary between adjacent cells is represented by a straight segment, the model does not appropriately express the curvature of the cell boundary. This can be overcome relatively easily, however, by adding vertices to split the straight segment into multiple segments.

If we try to extend the 2D cellular vertex model to deal with curved cell boundaries with multiple segments, the expression of frictional force can be a major problem. In the current vertex model, only the vertices of polygons experience friction forces from the surrounding objects [27]. This means that if a cell boundary is divided into some number of segments to express its curvature more smoothly, the friction forces on the cell boundary can change depending on the number of segments. For example, let us consider the case where we represent a straight boundary between cells A and B in two ways. One is that the cell boundary is represented by one segment specified by vertices 1 and 2. The other is that we add a new vertex 3 to the segment 12 and split the segment into two segments (i.e., the cell boundary is now represented by two segments, segment 13 and segment 32). Then, we consider the case where the cell boundary shifts slightly. In that motion, the latter representation of the cell boundary obviously experiences a larger friction force than the former representation if the friction coefficients are the same, because the latter representation has three vertices whereas the former representation has two vertices. Friction force on the cell boundary should not depend on discretization of the cell boundary, but instead, should satisfy a continuous limit (*i.e.,* the frictional force is expressed by a quantity per unit length of the cell boundary). Recently a 2D continuous mathematical model has been proposed that successfully reproduces the adhesion-independent movement of a single cell confined in a 3D space under axisymmetric conditions [32]. However, this model can assess only one cell. To comprehensively investigate cell migration, mechanical models capable of treating multiple cells that adhere to each other and satisfying force balance on the cell surface are necessary.

Here, we present a 2D cell membrane model, in which cells on a substrate are represented by polygons, and various forces on the cell surface, including contraction forces from actomyosin, adhesion between cells, and hydrostatic pressure from the cytosol, are implemented and balanced on the surface at all times. This model is equivalent to a continuous model if the lengths of segments representing the cell surface are kept within an appropriate range, using defined replacement rules, as described below. Cell clusters are represented by allowing a common surface between adjacent cells. We show that if cells in this model have a polarity expressed by direction-dependent surface tension, this causes the cell surface to flow from front to rear, and that flow induces unidirectional cell movement, not only for a single cell but also for a cluster of cells. In addition, if the surface-tension polarities of cells within the cluster are tilted with respect to the center of the cluster, the clusters rotate around the cluster center.

This mechanical model produces movement while keeping forces balanced on the cell surface due to the inflow and outflow of cell surface components from inside the cell. The relationship between cell migration speed and turnover rate of cell surface components is discussed in Section 3.5.

## 2 Model

### 2.1 The situations treated by the mechanical model

We consider situations where cells are on a substrate; some cells exist individually on the substrate, while other cells are attached to each other to form a cluster. We focus on the dynamics of the peripheries of the cells, which movements are determined by force balance on the cell boundaries. Each cell has a polarity, and depending on the direction of the polarity, each cell changes the strength of contraction force on the cell boundary. We are concerned with two setups of cell polarity. One is that there is a chemoattractant gradient in a definite direction on the substrate, say the *x*-direction, and all cells have polarity in the *x* direction (Sections 3.2-3.6). The other is that cells that form a cluster have different directions of polarity. To be specific, the direction of polarity of each cell tilts with respect to the center of the cluster (Section 3.7). This tilted polarity may be achieved by chance or by using chirality of the cells [33-35]. A characteristic point of our model is that force balance holds at any parts of the cell boundaries at any time.

### 2.2 Setup of the model

In our model, cells on a substrate are represented by polygons; specifically, the $\alpha$-th cell is represented by a polygon that consists of $N_\alpha$ segments and $N_\alpha$ nodes (Figure 1A). The number $N_\alpha$ can differ from cell to cell and changes with time, based on the following rules: if the segment under consideration becomes longer than some critical length $\ell^*_{long}$, it is split into two segments by creating a new node at the center of the segment, and if the segment under consideration becomes shorter than some critical length $\ell^*_{short}$, it is replaced by a single node whose position is the center of the segment before the replacement (Figure 1B). Using appropriate $\ell^*_{long}$ and $\ell^*_{short}$ values, the surface of the cell becomes smooth enough, and the dynamics of this discretized model are consistent with those observed in continuous models, as shown in Figures 1C and 2B, C. When a cell adheres to an adjacent cell, the segments and nodes on the cell boundary are shared between the two cells; that is, they have the same segments and nodes at their boundary (Figure 1A). Quantities assigned to the cell boundary, such as surface tension and rigidity of the cell boundary, are obtained by considering each quantity on the surfaces of the two cells, as in the existing vertex models [27].

Cell boundaries experience two types of forces: one is the frictional forces that are expressed by the dissipation function given in Equation (1), and the other is the mechanical forces that are expressed by a potential function $U$ in Equation (2). Frictional forces arise from both external and internal factors. When the segments comprising the cells move with respect to the substrate, each segment bears a frictional force from the substrate, whose strength is proportional to the length and velocity of the segment. The frictional coefficient can depend on the direction of movement relative to the direction of the segment; that is, when the segment moves parallel to the direction of the segment, the frictional coefficient is $\eta_\parallel$, whereas when the segment moves perpendicular to the direction of the segment, the frictional coefficient is $\eta_\perp$. In addition, it is assumed that each segment has an internal friction; when a segment shrinks or expands, a resistance force arises within the segment. The strength of the internal friction is proportional to the strain rate of the segment, with the friction coefficient $\mu$. Elongation or shrinkage of the segment is assumed to be an affine transformation (*i.e.*, the segment is homogeneously elongated or shrunken). Under these assumptions, we can calculate the dissipation function $W$, which gives the frictional forces on segments in terms of only the positions and velocities of nodes on the cell surfaces (see Appendix A for details), as

$$W = \sum_{\langle ij \rangle} \left\{ \frac{\eta_\parallel^{(ij)} \ell_{ij}}{2} \left( (\dot{X}_{ij} \cos\theta_{ij} + \dot{Y}_{ij} \sin\theta_{ij})^2 + \frac{(\dot{\ell}_{ij})^2}{12} \right) \right. $$
$$\left. + \frac{\eta_\perp^{(ij)} \ell_{ij}}{2} \left[ \left( \dot{X}_{ij}^2 + \dot{Y}_{ij}^2 + \frac{(\dot{\ell}_{ij})^2}{12} + \frac{(\dot{\theta}_{ij} \ell_{ij})^2}{12} \right) - \left( (\dot{X}_{ij} \cos\theta_{ij} + \dot{Y}_{ij} \sin\theta_{ij})^2 + \frac{(\dot{\ell}_{ij})^2}{12} \right) \right] + \frac{\mu^{(ij)}}{2} \frac{(\dot{\ell}_{ij})^2}{\ell_{ij}} \right\}$$

. Equation (1)

Here, $\ell_{ij}$ is the length of the segment that connects the *i*-th and *j*-th nodes; that is, $\ell_{ij} = |\mathbf{r}^{(i)} - \mathbf{r}^{(j)}|$, where $\mathbf{r}^{(i)} = (x_i, y_i)$ is the position of the *i*-th node. $\mathbf{R}_{ij} = (X_{ij}, Y_{ij})$ is the

position of the center of the segment $ij$, given by $\mathbf{R}_{ij} = (\mathbf{r}^{(i)} + \mathbf{r}^{(j)})/2$, and $\theta_{ij}$ is the angle between the $x$-axis and the vector from the $i$-th node to the $j$-th node. The dot over a quantity represents its time derivative, and the symbol $\langle ij \rangle$ under the summation symbol means that the sum is taken over all the segments in the system. Note that the sets of $\{\mathbf{R}_{ij}\}$, $\{\theta_{ij}\}$ and $\{\ell_{ij}\}$ are expressed by the set of positions of nodes, $\{\mathbf{r}^{(i)}\}$. Thus, as mentioned above, $W$ is a function of $\{\mathbf{r}^{(i)}\}$ and $\{\dot{\mathbf{r}}^{(i)}\}$.

The frictional coefficients $\eta_\parallel^{(ij)}$, $\eta_\perp^{(ij)}$, and $\mu^{(ij)}$ in Equation (1) can vary from segment to segment. However, our main objective is to show some simple applications of this model. We, therefore, kept these coefficients the same for every segment, such that $\eta_\parallel^{(ij)} = \eta_\perp^{(ij)} = \eta$ and $\mu^{(ij)} = \mu$ throughout this paper. Differentiating $W$ in Equation (1) with respect to the velocity the $i$-th node gives the frictional force $\mathbf{F}_{friction}^{(i)}$ on the $i$-th node as $\mathbf{F}_{friction}^{(i)} = -\partial W / \partial \dot{\mathbf{r}}^{(i)}$ [36]. Note that the dissipation function $W$ in Equation (1) considers the length dependence of the frictional force on the segment. Thus, even if a segment $ij$ is divided into two segments, $ik$ and $kj$, by adding a new node $k$, the total frictional forces on the segments $ik$ and $kj$ are the same as those on the segment $ij$ before the division, provided movement of segments $ik$ and $kj$ is the same as that of segment $ij$. In this sense, the frictional force on the cell surface does not depend on the number of nodes on the cell surface and this model satisfies some continuous limits on the frictional forces.

The other mechanical forces on the cell surface, such as contraction forces coming from the actomyosin network beneath the plasma membrane and hydrostatic pressure from the cytosol, are represented by the following effective potential $U$, as follows:

$$U = \frac{K}{2}\sum_\alpha (A_\alpha - A_\alpha^{(0)})^2 + \frac{K_p}{2}\sum_\alpha (L_\alpha - L_\alpha^{(0)})^2 + \frac{1}{2}\sum_{\langle ijk \rangle} \kappa^{(ijk)} \left(\frac{\mathbf{r}^{(j)} - \mathbf{r}^{(i)}}{\ell_{ij}} - \frac{\mathbf{r}^{(k)} - \mathbf{r}^{(j)}}{\ell_{jk}}\right)^2 \frac{2}{\ell_{ij} + \ell_{jk}},$$
$$+ \sum_{\langle ij \rangle} \gamma_{ij}(t)\ell_{ij}$$

Equation (2)

where $K$, $K_p$, and $\kappa^{(ijk)}$ are non-negative constants. This form of $U$ is quite similar to that given in existing vertex models [29,27,37], although it differs in the introduction of the third term, which represents the bending energy of cell membranes, and the form of $\gamma_{ij}(t)$, which expresses the strengh of surface tension on the cell membrane and is related to cell polarity. The first term in Equation (2) represents a pressure acting on the segments arising from the area difference between the current area of the cell, $A_\alpha$, and its preferred constant value, $A_\alpha^{(0)}$. This pressure can be interpreted as the hydrostatic pressure in the cytosol on the membrane of the cell. The $\alpha$ under the summation symbol indicates that the sum is taken over all cells in the system. The second term in Equation (2) expresses the property that the cell tends to keep the perimeter length $L_\alpha$ at some preferable length $L_\alpha^{(0)}$. This term represents the tendency to conserve the amount of cell membrane [27,38]. The third term in Equation (2) expresses the bending energy of the membrane. The strength of the membrane against bending is characterized by the coefficient $\kappa^{(ijk)}$, such that, if $\kappa^{(ijk)}$ is large, the part of the membrane under consideration, expressed by segments $ij$ and $jk$, is difficult to bend, and *vice versa*. The symbol $\langle ijk \rangle$ under the summation symbol indicates that the sum is taken over all pairs of segments that are connected and adjacent to each other. Lastly, the final term in Equation (2) expresses the surface tension acting on the membrane, which is the result of both contraction forces from the actomyosin networks beneath the membrane and adhesion of the cell to outside objects, such as adjacent cells or substrate. Surface tension strength, expressed by $\gamma_{ij}$, is controlled by the cell via altered expression of proteins, such as myosin, actin, cadherin, and integrin. When contraction force is strong or adhesion is weak at segment $ij$, $\gamma_{ij}$ becomes large, whereas when adhesion at segment $ij$ is strong or contraction force is weak at segment $ij$, $\gamma_{ij}$ becomes small.

Using this value of $\gamma_{ij}$, we can express the polarity of the cell. For example, consider the case where a cell has a polarity in the *x*-direction, and assume that the cell has a weak contraction force at the leading edge ($x > x_0$, where $x_0$ is the *x*-component of the position of the cell center) and a strong contraction force at the rear ($x < x_0$). This situation is expressed by letting the value of $\gamma_{ij}$ depend on the relative position of the cell membrane with respect to the center of the cell. That is, a small value is assigned to $\gamma_{ij}$ at the front of the cell, and a large value is assigned to $\gamma_{km}$ at the rear. In this paper, we consider several cases with different assigned $\gamma_{ij}$ values. If we accept the form of $U$ given in Equation (2), the total mechanical force $\mathbf{F}_U^{(i)}$ from $U$ on the *i*-th node is given by $\mathbf{F}_U^{(i)} = -\partial U / \partial \mathbf{r}^{(i)}$.

In this model, we assume that the inertial force of the cell surface is negligible compared with the mechanical forces in question, and thus, all forces acting on the *i*-th node coming from $U$ and $W$ must be balanced at all times. That is,

$$-\frac{\partial W}{\partial \dot{\mathbf{r}}^{(i)}} - \frac{\partial U}{\partial \mathbf{r}^{(i)}}\bigg|_{\gamma_{jk} = \hat{\gamma}_{jk}} = 0$$

Equation (3)

holds for all *i*'s at any given time. Here, the symbol $\gamma_{jk} = \hat{\gamma}_{jk}$ in the second term indicates that, after the derivative with respect to $\mathbf{r}^{(i)}$, each $\gamma_{jk}$ in the second term is replaced by an explicit value $\hat{\gamma}_{jk}$, which is expressed as a function of $\{\mathbf{r}^{(i)}\}$ [27,37,38]. This operation means that $\gamma_{jk}$ obeys some other dynamics that are much faster than those of $\{\mathbf{r}^{(i)}\}$, such that $\gamma_{jk}$ immediately reaches a value determined by $\{\mathbf{r}^{(i)}\}$ [38]. If we accept that the dynamics of localization of molecules related to contraction, such as myosin, on the cell surface (tens of seconds) is much faster than that of cell membrane movement (a few minutes), this operation may be allowed. Equation (3) gives the time evolution equations for $\{\mathbf{r}^{(i)}\}$.

## 2.3 Methods for numerical simulations

To numerically solve Equation (3), we first nondimensionalize the variables and parameters appearing in the equation, using the following units: length, $\sqrt{A_0/\pi}$; time, $\eta/(K_0\sqrt{A_0/\pi})$; and energy, $K_0(A_0/\pi)^2$; where $K_0$ is a typical value of $K$ solving for $U$ in Equation (2). Hereafter, we imply that the variables and parameters appearing in our model are nondimensionalized with the above units. For example, $A_0 = \pi$ and $\eta = 1$. We then numerically solve the nondimensionalized Equation (3) for $\{\mathbf{r}^{(i)}\}$ by the Euler method, with the step size dt = 1/5000 or 1/10000. For each step, we further determine whether the length of each segment satisfies the replacement criteria $\ell_{long}^* < \ell_{ij}$ or $\ell_{short}^* > \ell_{ij}$. Any segments that satisfy either of these two conditions are replaced by the rules shown in Figure 1B.

## 3 Results

### 3.1 A case in which a circular cell isotropically shrinks with a constant surface tension

To test our model, we first determine whether the dissipation function $W$ given in Equation (1) appropriately expresses the frictional forces on the cell surface. For this we consider a case where a circular cell has a constant surface tension $\gamma = 1$ and no constraints on its area $A$ and perimeter $L$; that is, we set $K = 0$, $K_p = 0$, and $\kappa^{(ijk)} = 0$ in Equation (2). In this case, the circular cell shrinks with some speed, while maintaining its circular shape. The time series of the radius of the circular cell, $r(t)$, is analytically obtained (Appendix B), resulting in $r(t) = \sqrt{\dfrac{\mu}{\eta} W_0(\dfrac{\eta}{\mu} e^{-\dfrac{2}{\mu}(\gamma_0 t - (\dfrac{\eta}{2} r_0^2 + \mu \log r_0))})}$, where $W_0(x)$ is the Lambert W function (the product logarithm) that satisfies $x = W_0(x) e^{W_0(x)}$ for $x > -1/e$. Comparing this analytical solution with the results of numerical simulation for Equation (3) reveals good agreement (Figure 1C). Thus, we conclude that the friction force expressed by Equation (1) appropriately expresses the frictional force experienced by the cell membrane, at least for this simple test case.

## 3.2 A circular cell migrates due to direction-dependent surface tension

Next, we investigate a more realistic case wherein the cell under consideration is circular with a constant area $A = A_0$ and has a polarity in the *x*-direction. Constant area of the cell is achieved by using the parameters $K \gg 1$, $K_p = 0$, and $\kappa^{(ijk)} = 0$ in Equation (2), and cell polarity is expressed by the direction-dependent surface tension $\hat{\gamma}_{ij}$ in Equation (3) as

$$\hat{\gamma}_{ij} = \gamma_0 - a \cos \theta_{ij}^{(\alpha)},$$

**Equation (4)**

where $\gamma_0$ is a positive constant, $a$ is a non-negative constant, and $\theta_{ij}^{(\alpha)}$ is the angle between the *x*-axis and the vector connecting the center of cell $\alpha$ that contains segment $ij$ and the center of segment $ij$ (see Figure 1A). The value of $a$ represents the degree of the polarity, such that, when $a = 0$, the cell is isotropic and has no polarity, whereas when $a > 0$, the cell boundary at a relatively large x has a weak surface tension, and the cell boundary at a relatively small x has a strong surface tension. Hereafter, we refer to the region of cell boundary at a relatively larger x as the "front" of the cell, and the region of the cell boundary at a relatively smaller x as the "rear" of the cell. If $a = 0$, the shape of the cell is circular due to the constant area and constant surface tension, $\gamma_0$, on the cell surface. In this simulation, we set $a$ as a small positive value ($a \ll \gamma_0$), such that the circular shape of the cell is still retained.

We then examined the steady state of the cell with these model parameters. Numerical simulations reveal that the cell moves in the positive *x*-direction, with a constant speed, while keeping a circular shape in the steady state (Figure 2A). The driving force for this movement is the direction-dependent surface tension in Equation (4) and the resulting cell surface flow from the front to the rear (Movie S1). That is, in the front region of the cell, surface tension is relatively weak due to the parameters of Equation (4), such that the surface in the front region tends to expand. In contrast, surface tension in the rear region of the cell is relatively strong, and hence, the surface in this region tends to shrink. This surface tension-dependent tendency of the cell surface to expand or shrink produces a flow of cell membrane, in which the front region is always expanding (in some sense, blebbing is continuously occurring in front of the cell), whereas the rear region is always shrinking (Movie S1). These behaviors act as a source and sink of cell surface and yield a flow of cell surface from front to rear. In addition, within this system, there is a frictional force between the cell surface and the substrate. Thus, if the cell surface moves in some direction, the whole cell experiences forces that move it in the opposite direction, based on the action–reaction principle.

The velocity $\mathbf{V}$ of this cell movement resulting from the direction-dependent surface tension can be analytically calculated, by treating the cell as a continuous circular object (see Appendix C for details). The result is

$$\mathbf{V} = (a/(\eta R(1+2\mu/(\eta R^2))), 0),$$

**Equation (5)**

where $R$ is the radius of the circular cell. We then compared this analytical solution with the cell speed obtained by numerically solving Equation (3) and found that these values are in good agreement with each other (Figure 2B, C). This indicates that our discrete model described by Equations (1)–(3) can describe continuous cell surface dynamics if we choose appropriate values for parameters, such as $\ell^*_{long}$ and $\ell^*_{short}$.

### 3.3 An elongated cell migrates due to direction-dependent surface tension

We further find that single-cell migration induced by direction-dependent surface tension, described in Equation (4), occurs even when the cell shape is elliptical (Figure 2D). We can model an elliptical-shaped cell using finite values for $K_p$ and $\kappa^{(ijk)}$ in Equation (2) and by ensuring the preferred cell perimeter $L_0$ is longer than that of a circle with the area $A_0$, $2\sqrt{\pi A_0}$. Overall, the mechanism for cell movement is basically the same as for circular cells, that is, the surface flows from front to rear due to direction-dependent surface tension, described in Equation (4). In this simulation, we set the initial configuration of the cell, such that the long axis is aligned in the y-direction (Figure 2D; $t$=0). Initially, up until $t$=8, the elliptical cell moves in the x-direction, while keeping the short axis of the cell aligned with the x-axis. However, at around $t$=25, the long body axis begins to incline toward the x-direction, and the direction of cell movement also begins to incline; that is, the cell has the y-component of velocity (Movie S2). In the final stage, the long body axis is completely oriented to the x-axis, and the cell moves in the x-axis again. The final speed of the cell is faster than in the earlier stage, where the long axis of the cell is perpendicular to the x-axis. This simulation indicates that if the cell shape is elliptical and the direction of cell polarity is fixed in the *x*-axis, cell movement in which the short axis is oriented to the *x*-axis is slow and unstable, whereas cell movement in which the long body axis is oriented to the *x*-axis is fast and stable. In general, the parameters of Equation (4) make cell movement fast if the long axis of the cell is directed in the direction of cell polarity. This property clearly appears in the next case, where the cell is sandwiched by two parallel walls.

### 3.4 A case in which a cell is sandwiched by two walls

Motivated by results from published experiments [19,39,40], we next examined the scenario in which a cell is sandwiched by two parallel walls. In this case, the wall is expressed using the potential $U_{wall}(y) = (y-d)\Theta(y-d) - (y+d)\Theta(-y-d)$, where $d$ is one-half the distance between the walls, and $\Theta(x)$ is the Heaviside step function, defined as $\Theta(x) = 1$, for $x > 0$, and $\Theta(x) = 0$, for $x \leq 0$. By adding the terms $\sum_{all\ nodes\ i's} U_{wall}(y^{(i)})$ to Equation (2), the cell surface close to the walls experiences a repulsive force from the wall potential. In this setup, the area of the cell is held constant as $A = A_0$, which is achieved with the same parameters as used for the simulation with a circular cell. Under these conditions, the cell also migrates in the *x*-direction in the steady state (Figure 3A), where the speed of the cell increases as $d$ decreases (Figure 3B), consistent with published experimental results [39,40]. To demonstrate how and why cell speed increases with decreased $d$, we derive the analytical expression for the speed of the cell (see Appendix C), which is given by

$$V_x = \frac{2}{\pi \eta d}\left(\int_0^{\pi/2} \sin\xi_1 \frac{d\gamma(\theta_{forward}(\xi_1))}{d\xi_1} d\xi_1 + \int_{\pi/2}^{\pi} \sin\xi_2 \frac{d\gamma(\theta_{backword}(\xi_2))}{d\xi_2} d\xi_2 \right) + \frac{2\Delta\gamma}{\pi\eta d},$$

Equation (6)

where $\theta_{forward}(\xi_1) = \arctan(\frac{d\sin\xi_1}{d\cos\xi_1 + L_x})$, $\theta_{backword}(\xi_2) = \arctan(\frac{d\sin\xi_2}{d\cos\xi_2 - L_x})$, and $\Delta\gamma = \gamma(\pi - \theta_1) - \gamma(\theta_1) = 2aL_x/\sqrt{L_x^2 + d^2}$. The angles $\theta_{forward}$, $\theta_{backward}$, $\xi_1$, and $\xi_2$ used here are defined as shown in Figure 3C. The values $L_x$ and $d$ are the half-length of the straight part of the cell and the radius of the circular part of the cell, respectively. Because the total area of the cell is kept at $A = A_0$, $L_x$ and $d$ are related as $4L_x d + \pi d^2 = A_0$.

Results from analytical expression (6) are in good agreement with the results of numerical simulations (Figure 3B). Thus, we use Equation (6) to interpret the cell speed in this scenario. The first term in Equation (6) is the contribution from surface flow on the two semicircles of the cell, and the second term is the contribution from surface flow on the straight parts of the cell. When $d$ becomes small, the semicircles also become small, and the flow speed of cell surface along the semicircles becomes slow due to the surface tension gradient along the small semicircles. Thus, the first term in Equation (6) becomes small when $d$ decreases (Figure 3B, yellow dashed curve). In contrast, $\Delta\gamma$ in the second term in Equation (6), which is the surface tension difference between the two edges of the straight parts of the cell (Figure 3C), increases with decreased $d$ (see Equation (4)). Furthermore, $d$ in the denominator of the second term in Equation (6), which represents the resistance of the semicircle parts to the cell movement, also enlarges this term as the value of $d$ decreases. Thus, the whole second term in Equation (6) drastically increases when $d$ decreases (Figure 3B, green dashed curve). The tendency that the migration speed $v$ of the sandwiched cell increases with decrease in $d$ appears even when we take a different setup of $\hat{\gamma}_{ij}$. For example, we consider the case where the surface tension is constantly increased along the cell surface form front to rear, in which the minimum and maximum surface tensions are kept to be constant, i.e., $\hat{\gamma}_{ij} = 0.8$ at the front and $\hat{\gamma}_{ij} = 1.2$ at the rear. These values are the same as those in the simulations in Figure 3B. The results of numerical simulations with this setup is given in Figure S2, where as in Figure 3B $v$ increases with decrease in $d$, while the $d$ dependence of $v$ is somewhat moderate compared to Figure 3B. This same tendency of $v$ implies that the increase in $v$ of the sandwiched cell with decrease in $d$ may be a robust phenomena, as frequently observed in experiments [19, 40].

### 3.5 Properties of surface flow-mediated cell migration

To better understand why and how cells move due to surface flow while keeping the forces balanced at all times, we surveyed the properties of cell movement observed in this model. Although these properties derived below are provided in a continuous form, it is not difficult to translate them into a discrete form. As noted above, our model neglects all inertia of the cell surface, so that the forces on any element of the cell surface are balanced at any time, and more importantly, the total force on the cell surface from the substrate must vanish at any time. Because the force on the cell surface from the external object—the substrate—is friction only, we have the following equality:

$$\int_\Omega \eta \mathbf{v}(\xi,t) \frac{\partial s(\xi,t)}{\partial \xi} d\xi = 0,$$

Equation (7)

at any time $t$. Here, $\xi$ is the material coordinate that is assigned to the element of cell surface under consideration and $s(\xi,t)$ is the counter length of the arc of cell surface between some reference point on the cell surface and the cell surface element specified by $\xi$ at time $t$. Additionally, $\mathbf{v}(\xi,t)$ is the velocity of the surface element at $\xi$ and $t$; $\Omega$ under the integral symbol indicates that the range of the integral is the whole cell surface. We then consider the situation where cell movement reaches a steady state, with cell velocity and cell shape constant in time, focusing on some time $t_0$ in this steady state. At $t = t_0$, we reassign the material coordinate $\xi$ on the cell surface, such that $\xi$ coincides with the counter length $s$; that is,

$$s(\xi, t_0) = \xi.$$

Equation (8)

With this situation, let us consider the time evolution of the cell surface density, $\rho$. In general, $\rho$ obeys the following time evolution equation

$$\partial \rho(\xi,t)/\partial t = -((\partial \mathbf{r}/\partial \xi)\cdot(\partial \mathbf{v}/\partial \xi)/(\partial s/\partial \xi)^2)\rho + J(\xi,t),$$

Equation (9)

Where $\mathbf{r}(\xi,t)$ is the position vector of the surface element at $\xi$ and $t$, and $J$ is the flux of cell surface component from the cell inside to the cell surface per unit length. The derivation of Equation (9) is shown in Appendix D. In the steady state, $\rho$ is constant in time, say $\rho = \rho_0$, and the time derivative $\partial \rho/\partial t$ is zero. Thus, from Equations (8) and (9) we obtain

$$J(\xi, t_0) = \rho_0 (\partial \mathbf{r}/\partial \xi)\cdot(\partial \mathbf{v}/\partial \xi),$$

Equation (9.1)

which describes the relationship between $J$, $\mathbf{r}$, and $\mathbf{v}$ in the steady state. Because of the constant cell shape in the steady state, the total flux of the cell surface components must vanish at any time. Thus,

$$\int_\Omega J(\xi, t_0) d\xi = 0.$$

Equation (10)

Interestingly, $J$ in Equation (9.1) is related to the cell migration velocity $\mathbf{V}$ in the steady state for the following reasons. In the steady state, the cell surface center of mass also moves with the same velocity $\mathbf{V}$. However, as indicated in Equations (7) and (8), there is no net velocity of components on the cell surface, i.e., $\int_\Omega \mathbf{v}(\xi,t_0)d\xi = 0$. This implies that the shift in cell surface center of mass is, in fact, achieved by the inflow and outflow of cell surface components from inside the cell. Thus, the integral of the product of $J(\xi,t_0)$ and $\mathbf{r}(\xi,t_0)$ over the whole cell surface gives the transport rate of cell surface components through the inside of the cell. Dividing this quantity by the total amount of components on the cell surface, $\rho_0 L_{cell}$, yields the velocity of the cell surface center of mass. Therefore,

$$\mathbf{V} = \int_\Omega \mathbf{r}(\xi,t_0)J(\xi,t_0)d\xi / \rho_0 L_{cell},$$

Equation (11)

where $L_{cell}$ is the cell perimeter in the steady state. Equation (11) is interesting because this relation connects apparently different quantities, the cell migration velocity $\mathbf{V}$ and the turnover rate $J$ of the cell surface. Indeed, we can confirm that Equations (7), (10), and (11) exactly hold for the circular cell migration case with $\mu = 0$ as shown below. From the analytical results given in Equation (C.16) in Appendix C, $\mathbf{r}$ and $\mathbf{v}$ of the circular cell in the steady state are given as

$$\mathbf{r}(\xi) = (\frac{a}{\eta R}t + R\cos\frac{\xi}{R}, R\sin\frac{\xi}{R})$$
$$\mathbf{v}(\xi) = (\frac{a}{\eta R} - \frac{2a}{\eta R}(\sin\frac{\xi}{R})^2, \frac{2a}{\eta R}\sin\frac{\xi}{R}\cos\frac{\xi}{R}).$$

Equation (12)

From Equations (9.1) and (12), we have

$$J = \rho_0 \frac{2a}{\eta R^2}\cos\frac{\xi}{R}.$$

Equation (12.1)

Inserting Equations (12) and (12.1) into Equation (11) gives

$$\mathbf{V} = (\frac{a}{\eta R}, 0),$$

Equation (12.2)

where we have used $L_{cell} = 2\pi R$ and the fact that the range of $\xi$ is $[0, 2\pi R]$. Equation

(12.2) coincides with Equation (5) for the case of $\mu = 0$.

### 3.6 Multiple cells in a cluster also move via direction-dependent surface tension

The above simulations focus on migration of individual cells; however, cells often move together with other cells. Therefore, we next examined whether cell migration due to direction-dependent surface tension occurs even when multiple cells are attached to each other, forming a cluster. Specifically, we investigated the case where the number $N_{cell}$ of cells on the substrate is $N_{cell} = 2, 3, 4, 10$ with some initial configuration in which the cells are attached to each other (Figure 4C). In these simulations, we have set the surface tension on the boundary between the cells to a constant value (*i.e.*, $\gamma = \gamma_b$), by assuming that the same type of cells is adhering to each other with some characteristic strength (Figure 4A). The other cell boundaries, which do not contact other cells, have the direction-dependent surface tension specified by Equation (4). As in the case of single-cell migration investigated in Sections 3.2-3.4, each cell in this system keeps its area constant and has no constraint on its perimeter; that is, $K = 100$, $K_p = 0$, and $\kappa^{(ijk)} = 0$ in Equation (2). Under these conditions, multiple cells move in the *x*-direction, while maintaining attachment between cells in the steady state (Figure 4C, Movies S3–S10). Moreover, the mechanism of multiple-cell migration is basically the same as that of single-cell migration. That is, due to the direction-dependent surface tension of each cell, the surface of the cell cluster flows from front to rear, and this flow drives movement of the whole cluster.

The shape and velocity of the cluster during this movement depend on the value of $\gamma_b$. When $\gamma_b$ decreases, the shape becomes round, and speed becomes slow (Figure 4B, C). The roundness of the cell cluster at small $\gamma_b$ results from the constant area and lack of constraint on the perimeter of each cell. Due to these parameters, when $\gamma_b$ becomes small, boundaries between cells tend to extend, and the whole cell cluster behaves like one object. Further, because the cell cluster surface has a constant surface tension, $\gamma_0$, from Equation (4), which reduces the cluster perimeter as much as possible under the constant area, $A_{cluster} = N_{cell} A_0$, the whole cluster is nearly a circle. In contrast, when $\gamma_b$ becomes large, boundaries between cells tend to shrink, and outer cells within the cluster tend to be round, with the surface tension $\gamma_0$ (Figure 4C).

The slow cluster movement speed at small $\gamma_b$ results from the roundness of the cell cluster. As shown in Sections 3.2-3.4, surface flow-induced cell migration, in general, becomes fast when the cell is elongated in the *x*-axis. This tendency is most evident in the case of a cell sandwiched by two walls, described in Section 3.4, where $\Delta\gamma$ in Equation (6) acts as the driving force for movement. Similarly, in the case of cluster migration, the surface tension gradient along the cell surface (and the resultant flow of cell surface) is also the driving force for movement. Thus, if cells within the cluster are relatively elongated in the *x*-direction, the speed of cluster movement is relatively fast. However, when the cluster is round, such as for $N_{cell} = 2, 4$ and $\gamma_b = 0.1$ (Figure 4C), each cell in the cluster is relatively elongated in the *y*-axis, and thus, the driving forces are small, and the movement becomes slow.

## 3.7 Cell clusters rotate when cell direction is tilted with respect to the center of a cluster

To this point, we have examined only scenarios in which all cells on a substrate have the same direction of polarity (*e.g.,* the *x*-direction). However, because clustered cells may also show distinct polarities, we next investigated the case wherein direction of polarity for individual cells within a cluster differs from cell to cell. In particular, we considered four cells comprising a cluster, in which each individual cell has a polarity that is directed perpendicular to the direction from the center of the cell to that of the cell cluster (Figure 5A). Surface tension for cells within this situation is expressed as

$$\hat{\gamma}_{ij} = \gamma_0 - a\cos(\theta_{ij}^{(\alpha,c)} - 3\pi/2),$$

Equation (13)

where $\theta_{ij}^{(\alpha,c)}$ is the angle between the vector from the center of cell $\alpha$ that contains segment $ij$ to the center of the cluster and the vector from the center of cell $\alpha$ to the center of segment $ij$ (Figure 5A). The other parameters for this simulation are the same as those for cluster migration investigated in Section 3.6; that is, boundaries between cells within the cluster have a constant surface tension, $\gamma_{ij} = \gamma_b$, and each cell in the cluster keeps its area constant and has no constraint on its perimeter, with $K = 100$, $K_p = 0$, and $\kappa^{(ijk)} = 0$ in Equation (2).

The initial configuration of the cell cluster is the same as in Figure 4C ($N_{cell} = 4$). Numerical simulations with the above parameters further indicate that each cell within the cluster continues to rotate counterclockwise around the center of the cluster while maintaining its attachments (Figure 5B, Movies S11 and S12). This rotation continues as long as each cell has the polarity established in Equation (13), and the mechanism of movement is basically the same as observed for cluster migration and single-cell migration. That is, cells move due to the direction-dependent surface tension expressed by Equation (13); the surface of the cell cluster moves clockwise, and this surface movement generates the driving force for rotational movement of the whole cluster. In this simulation, we set the polarities of the cells in the cluster with Equation (13). However, in reality it may be more reasonable that the direction of cell polarity changes with time by following some other rules. One such rule is called the velocity alignment mechanism [41], in which cells align their polarity to their velocity. We incorporated this rule into our model and performed numerical simulations. The results show that for any initial directions of cell polarity, the cell polarities finally align such that the cluster of cells rotates (Movies S13, S14). The direction of rotation depends on the initial distribution of cell polarities in the cluster. That is, the cell cluster rotates clockwise or counterclockwise depending on the initial distribution of cell polarities. Our system does not require cell confinement for this rotational motion, which is different from the results of previous studies [41].

As in the case of single-polarity cluster migration examined in Section 3.6, shape and rotational velocity of the cell cluster change with $\gamma_b$ (Figure 5B, C). When $\gamma_b$ is small, the cell cluster becomes round, and the angular velocity for the rotational movement of the cluster becomes large. The reasons for this cluster roundness are the same as those outlined in Section 3.6, that is, the whole cluster behaves like one object due to a lack of constraint on cell perimeter and small $\gamma_b$. In addition, the constant surface tension, $\gamma_0$, reduces cluster surface as much as possible under the constant whole area, $A_{cluster} = N_{cell} A_0$.

The faster rotation of the cell cluster at small $\gamma_b$ originates from the velocity distribution of the cell cluster surface (Figure 5D(a)), and the reasons for this are as follows. First, cell boundaries within the cluster are classified into one of two types: (i) cell boundaries that are located at the surface of the cluster and are associated with only one cell, referred to as the "outer cell boundaries" and (ii) cell boundaries that are located within the cluster and form boundaries between two adjacent cells, referred to as "inner cell boundaries." During rotation of the cluster, the driving forces for cluster movement come from the flow of the outer cell boundaries, because only the outer cell boundaries have cell polarity, as indicated by Equation (13). In the case of the round cluster ($\gamma_b = 0.2$; Figure 5D(a)), most velocities of the outer cell boundaries are directed in the tangent of the cluster surface, which comes from the monotonic decrease in surface tension along the surface. These velocities generate a large magnitude of torque compared with those present in the case of $\gamma_b = 1.0$ (see Figure 5D, E). Here, $N_z^{(flow)}$ is the z-component of the torque generated by the surface flow of the outer cell boundaries and is evaluated as $N_z^{(flow)} = -\eta \sum_{\langle ij \rangle_{outer}} ((\mathbf{R}_{ij} - \mathbf{r}_c) \times (\dot{\mathbf{R}}_{ij} - \dot{\mathbf{r}}_c))_z \ell_{ij}$, where $\mathbf{r}_c$ is the position vector of the cluster center, and $\langle ij \rangle_{outer}$ indicates that the summation range is over all outer cell boundaries. $N_z^{(flow)}$ is balanced with the z-component of torque generated from frictional forces experienced by the inner cell boundaries, denoted by $N_z^{(friction)}$, and this balance determines the angular velocity, $\omega$, of the cluster rotation. $N_z^{(friction)}$ is roughly expressed as $N_z^{(friction)} = -\zeta \omega$, where $\zeta$ is a coefficient given by $\zeta = \eta \sum_{\langle ij \rangle_{inner}} |\mathbf{R}_{ij} - \mathbf{r}_c|^2 \ell_{ij}$.

The $\zeta$ coefficient is almost constant in time because it is determined by the shape of the network of inner cell boundaries, which is also almost constant in time (see Movies S11, S12). The torque balance on the cell surface gives $\omega = N_z^{(flow)} / \zeta$. Although both $N_z^{(flow)}$ and $\zeta$ increase with decreased $\gamma_b$ (Supplementary Figure S1), the increase in $\zeta$ is more gentle than for $N_z^{(flow)}$, which results from the fact that the size of the network of inner cell boundaries does not change much with $\gamma_b$ (see Figure 5B). Thus, the increase in $N_z^{(flow)}$, which occurs with decreased $\gamma_b$, increases the angular velocity $\omega$.

## 4 Discussion

In this study, we have developed a 2D cell membrane model, described by Equations (1)–(3), which shows that cells on a substrate migrate due to direction-dependent surface tension, represented by Equation (4). Notably, this is true, not only for a single cell but also for multiple cells that adhere to one another and form a cluster. Moreover, if we change the direction of cell polarity within a cell cluster, such that polarity varies from cell to cell, as in Equation (13), the cell cluster rotates due to direction-dependent surface tension. A key point of emphasis is that, because this model neglects any inertia of the cell surface and ensures that forces are balanced on the cell surface at all times, the total force exerted on each cell from outside objects, such as the substrate and neighboring cells, always vanishes. That is, the cell movement in this situation is, what we call, "force free", and this is driven by the inflow and outflow of cell membrane components from inside the cell, as discussed in Section 3.5.

Although this is a 2D model, it is expected that the cell movement we observe herein, namely, single-cell migration, cluster migration, and cluster rotation, may also be present in 3D models, because the mechanism of movement (*i.e.*, direction-dependent surface tension) is likely applicable to 3D cases. Indeed, single-cell migration due to direction-dependent surface tension has been demonstrated in a previously published study [26], for a spherical cell migrating with a constant velocity in 3D space. Thus, we anticipate that cluster migration and rotation will be shown in well-implemented 3D models.

We further expect this model may reveal relevant information for better understanding cell shape and cortical flow during cell movement in real-life situations. This is because, as previously emphasized, our model satisfies force balance on the cell membrane at all times, and hence, the cell shapes appearing in this model are the results of that force balance on cell surface during cell migration (*e.g.,* see Figure 2D and Figure 4C). Thus, if a parameter set in our model mimics the cell shapes and cell movement observed in experiments, we can expect which part of the cell experiences strong contractile forces and which part of the cell membrane has strong stiffness by looking at the experimental results. In addition, this mechanical model provides the speed of cell movement. Thus, with these properties we can address the relationship between cell shape and cell velocity, which is proposed in the existing works [42] by using a symmetry argument. We can examine the relationship between shape and speed of cells by the cell-level model. In addition, by regarding the segments in our model as the cortex in cells, we are able to investigate cortical flow within moving cells, as in previously published studies [32]. A key point of the present model is that due to the discreteness of this model, we can easily set the parameter values that specify key characteristics of the cells we are modeling. For example, we can change the strength of contractile and frictional forces, as well as the stiffness of the cortex, by changing $\gamma_{ij}$, $\eta_{ij}$, and $\kappa^{(ijk)}$ in space and time. For example, cells change its behaviors depending on the stiffness of the substrate (Durotaxis) [43]. By setting up that the values of $\gamma_{ij}$, $\eta_{ij}$, and $\kappa^{(ijk)}$ depend on the stiffness of the substrate in some way in our model, we can investigate effects of the stiffness of the substrate on the cell movement from a subcellular level.

In our model, as already mentioned in Section 2.1, we focused on the dynamics of the cell peripheries. However, in many real-life cases, cells are attached to the substrate at some focal adhesions, and the two-dimensional flow of the cortex at the bottom region of the cells is also important for cell movement. Therefore, it would be a good strategy to extend the presented cortex model to a model that expresses the two-dimensional flow on the plane. This will be a target of future work. In addition, it has been observed that the plasma membrane and the cortex beneath the plasma membrane move separately during cell movement [44]. This observation implies that it is more realistic for a model of cell movement to describe the dynamics of two components, the plasma membrane and the cortex. This will also be a target of future research.

In the current remodeling rules for segments that represent cell boundaries, given in Section 2.2, the attachment and detachment of cells, which are important processes for considering more general movement of cells, are not described. In fact, by adding only some rules to the current setup, we can describe the attachment and detachment of cells. An example of it is as follows. When two separated cells are close to each other and the distance between some parts of two cell surfaces becomes shorter than some critical distance, then the parts of cell surfaces are merged and becomes one segment. This setup describes the attachment of two cells. In addition, when the common cell boundary between two adhered cells becomes shorter than some critical length, then the common cell boundary is split into two cell boundaries and two cell are separated. This situation represents the detachment of two cells. By introducing this rule into the current version, we can investigate more complex behaviors of cells on a substrate. In the present paper, we have assumed that cell polarity is represented by the direction dependence of $\gamma_{ij}$. However, because the front and rear of a cell differ in regard to the stiffness of the cell surface and the strength of attachment to the substrate, it may be more plausible to model the situation in which cell polarity can change the cell stiffness, $\kappa^{(ijk)}$, as well as the friction coefficients, $\eta_{ij}$ and $\mu_{ij}$. By doing so, we may reproduce more realistic cell behaviors, such as blebbing [18]. We can also consider many variations of the form of $\hat{\gamma}_{ij}$. For example, since the position of nucleus in a cell plays an important role for cell dynamics [45], it may be more realistic thet the value of $\hat{\gamma}_{ij}$ is determined by the relative position of the cortex to the nucleus.

In the present paper cell polarity was a priori given as in Equations (4) and (13). But in reality, the direction of cell polarity dynamically changes with time depending on the environment surrounding the cell and the state of itself. Many possible ways for describing the time evolution of cell polarity have been proposed [41]. By introducing proposed descriptions of cell polarity into our mechanical model and comparing the results obtained by the model to experiments, we might be able to determine which time evolution rule for cell polarity is most appropriate.

The phenomenon of cells migrating on a substrate, while forming a cluster, has been observed during development of the zebrafish lateral line [7,46] and by the amoeba *Dictyostelium discoideum* [47,48]. In addition, some clusters of epithelial cells were found to rotate 90º within an epithelial sheet during *Drosophila* eye development [10,49,50,51]. Although there are many possible explanations for such collective behavior [7,46], it has been difficult to understand from a mechanistic standpoint, due to the many-body problem inherent in analyzing complex systems. In this regard, mechanical models are useful to better understand the forces that drive multi-cell behavior as noted in the Introduction, and the mechanism presented in this paper represents one possible explanation for such collective cell movement. By comparing experimental results, such as the spatial distribution of actomyosin and adhesion molecules in cells and observed cell shapes, with the results from our numerical simulations, we may uncover new insights into the complex patterns of cell movement observed in living systems.

## 5 Author Contributions

K.S. designed the mechanical model and performed analytical calculations and numerical simulations of this model. K.S. wrote the paper.

# 6 Acknowledgments

We would like to thank T. Nakagaki, H. Orihara, N. Nishigami, M. Nishikawa, M. Akiyama, M. Kimura, and Y. Tanaka for their valuable comments and discussions. This work was supported by a JSPS KAKENHI Grant, Number JP20K03871 and 21H05310, the "Dynamic Alliance for Open Innovation Bridging Human, Environment and Materials" of the Ministry of Education, Culture, Sports, Science and Technology of Japan, and the Global Station for Soft Matter at Hokkaido University.

# 7 References

1. Friedl P, and Gilmour D. Collective Cell Migration in Morphogenesis, Regeneration and Cancer. *Nat Rev Mol Cel Biol* (2009) 10(7):445-57.　doi:10.1038/nrm2720
2. Scarpa E, Mayor R. Collective cell migration in development. *J Cell Biol.* (2016) Jan 18;212(2):143-55. doi: 10.1083/jcb.201508047
3. Bodor DL, Pönisch W, Endres RG, Paluch EK. Of Cell Shapes and Motion: The Physical Basis of Animal Cell Migration. *Dev Cell.* (2020) Mar 9;52(5):550-562. doi: 10.1016/j.devcel.2020.02.013
4. Weijer CJ. Collective cell migration in development. *J Cell Sci.* (2009) Sep 15;122(Pt 18):3215-23. doi: 10.1242/jcs.036517
5. Montell DJ, Yoon WH, Starz-Gaiano M. Group choreography: mechanisms orchestrating the collective movement of border cells. *Nat Rev Mol Cell Biol.* (2012) Oct;13(10):631-45. doi: 10.1038/nrm3433
6. Haigo SL, Bilder D. Global tissue revolutions in a morphogenetic movement controlling elongation. *Science* (2011) Feb 25;331(6020):1071-4. doi: 10.1126/science.1199424
7. Haas P, Gilmour D. Chemokine signaling mediates self-organizing tissue migration in the zebrafish lateral line. *Dev Cell.* (2006) May;10(5):673-80. doi: 10.1016/j.devcel.2006.02.019
8. Pagès DL, Dornier E, de Seze J, Gontran E, Maitra A, Maciejewski A, Wang L, Luan R, Cartry J, Canet-Jourdan C, Raingeaud J, Lemahieu G, Lebel M, Ducreux M, Gelli M, Scoazec JY, Coppey M, Voituriez R, Piel M, Jaulin F. Cell clusters adopt a collective amoeboid mode of migration in confined nonadhesive environments. *Sci Adv.* (2022) Sep 30;8(39):eabp8416. doi: 10.1126/sciadv.abp8416

9. Kuwayama H, Ishida S. Biological soliton in multicellular movement. *Sci Rep.* (2013);3:2272. doi: 10.1038/srep02272
10. Founounou N, Farhadifar R, Collu GM, Weber U, Shelley MJ, Mlodzik M. Tissue fluidity mediated by adherens junction dynamics promotes planar cell polarity-driven ommatidial rotation. *Nat Commun.* (2021) Nov 30;12(1):6974. doi: 10.1038/s41467-021-27253-0
11. Callan-Jones AC, Voituriez R. Actin flows in cell migration: from locomotion and polarity to trajectories. *Curr Opin Cell Biol.* (2016) Feb;38:12-7. doi: 10.1016/j.ceb.2016.01.003
12. Innocenti M. New insights into the formation and the function of lamellipodia and ruffles in mesenchymal cell migration. *Cell Adh Migr.* (2018); 12(5):401-416. doi: 10.1080/19336918.2018.1448352
13. Case LB, Waterman CM. Integration of actin dynamics and cell adhesion by a three-dimensional, mechanosensitive molecular clutch. *Nat Cell Biol.* (2015) Aug;17(8):955-63. doi: 10.1038/ncb3191
14. Lämmermann T, Bader BL, Monkley SJ, Worbs T, Wedlich-Söldner R, Hirsch K, Keller M, Förster R, Critchley DR, Fässler R, Sixt M. Rapid leukocyte migration by integrin-independent flowing and squeezing. *Nature.* (2008) May 1;453(7191):51-5. doi: 10.1038/nature06887
15. Reversat A, Gaertner F, Merrin J, Stopp J, Tasciyan S, Aguilera J, de Vries I, Hauschild R, Hons M, Piel M, Callan-Jones A, Voituriez R, Sixt M. Cellular locomotion using environmental topography. *Nature.* (2020) Jun;582(7813):582-585. doi: 10.1038/s41586-020-2283-z
16. O'Neill PR, Castillo-Badillo JA, Meshik X, Kalyanaraman V, Melgarejo K, Gautam N. Membrane Flow Drives an Adhesion-Independent Amoeboid Cell Migration Mode. *Dev Cell.* (2018) Jul 2;46(1):9-22.e4. doi: 10.1016/j.devcel.2018.05.029
17. Farutin A, Étienne J, Misbah C, Recho P. Crawling in a Fluid. *Phys Rev Lett.* (2019) Sep 13;123(11):118101. doi: 10.1103/PhysRevLett.123.118101
18. Paluch EK, Raz E. The role and regulation of blebs in cell migration. *Curr Opin Cell Biol.* (2013) Oct;25(5):582-90. doi: 10.1016/j.ceb.2013.05.005
19. Bergert M, Erzberger A, Desai RA, Aspalter IM, Oates AC, Charras G, Salbreux G, Paluch EK. Force transmission during adhesion-independent migration. *Nat Cell Biol.* (2015) Apr;17(4):524-9. doi: 10.1038/ncb3134
20. Bergert M, Chandradoss SD, Desai RA, Paluch E. Cell mechanics control rapid transitions between blebs and lamellipodia during migration. *Proc Natl Acad Sci U S A.* (2012) Sep 4;109(36):14434-9. doi: 10.1073/pnas.1207968109

21. Ruprecht V, Wieser S, Callan-Jones A, Smutny M, Morita H, Sako K, Barone V, Ritsch-Marte M, Sixt M, Voituriez R, Heisenberg CP. Cortical contractility triggers a stochastic switch to fast amoeboid cell motility. *Cell.* (2015) Feb 12;160(4):673-685. doi: 10.1016/j.cell.2015.01.008
22. Bray D, White JG. Cortical flow in animal cells. *Science.* (1988) Feb 19;239(4842):883-8. doi: 10.1126/science.3277283
23. Salbreux G, Charras G, Paluch E. Actin cortex mechanics and cellular morphogenesis. *Trends Cell Biol.* (2012) Oct;22(10):536-45. doi: 10.1016/j.tcb.2012.07.001
24. Mietke A, Jülicher F, Sbalzarini IF. Self-organized shape dynamics of active surfaces. *Proc Natl Acad Sci U S A.* (2019) Jan 2;116(1):29-34. doi: 10.1073/pnas.1810896115
25. Bächer C, Khoromskaia D, Salbreux G and Gekle S (2021) A Three- Dimensional Numerical Model of an Active Cell Cortex in the Viscous Limit. *Front. Phys.* 9:753230. doi: 10.3389/fphy.2021.753230
26. Okuda, S., Sato, K. & Hiraiwa, T. Continuum modeling of non-conservative fluid membrane for simulating long-term cell dynamics. *Eur. Phys. J. E* (2022) 45, 69. doi:10.1140/epje/s10189-022-00223-0
27. Fletcher, A. G., Osterfield, M., Baker, R. E. & Shvartsman, S. Y. Vertex models of epithelial morphogenesis. *Biophys. J.* (2014) 106, 2291-2304.
28. Rauzi, M., Verant, P., Lecuit, T. & Lenne, P. F. Nature and anisotropy of cortical forces orienting Drosophila tissue morphogenesis. *Nat. Cell Biol.* (2008) 10, 1401-1410.
29. Farhadifar, R., Roper, J. C., Aigouy, B., Eaton, S. & Julicher, F. (2007) The influence of cell mechanics, cell-cell interactions, and proliferation on epithelial packing. *Curr. Biol.* 17, 2095- 2104.
30. Aigouy B, Farhadifar R, Staple DB, Sagner A, Röper JC, Jülicher F, Eaton S. Cell flow reorients the axis of planar polarity in the wing epithelium of Drosophila. *Cell.* (2010) Sep 3;142(5):773-86. doi: 10.1016/j.cell.2010.07.042
31. Sato, K., Hiraiwa, T., Maekawa, E., Isomura, A., Shibata, T. & Kuranaga, E. Left-right asymmetric cell intercalation drives directional collective cell movement in epithelial morphogenesis. *Nat. Commun.* (2015) 6, 10074.
32. Gaspard Jankowiak, Diane Peurichard, Anne Reversat, Christian Schmeiser and Michael Sixt. Modeling adhesion-independent cell migration. *Mathematical Models and Methods in Applied Sciences* Vol. 30, No. 03, pp. 513-537 (2020) doi: 10.1142/S021820252050013X
33. Wan LQ, Chin AS, Worley KE, Ray P. Cell chirality: emergence of asymmetry from

cell culture. Philos Trans R Soc Lond B Biol Sci. 2016 Dec 19;371(1710):20150413.
34. Taniguchi K, Maeda R, Ando T, Okumura T, Nakazawa N, Hatori R, Nakamura M, Hozumi S, Fujiwara H, Matsuno K. Chirality in planar cell shape contributes to left-right asymmetric epithelial morphogenesis. Science. 2011 Jul 15;333(6040):339-41. doi: 10.1126/science.1200940. PMID: 21764746.
35. Tee YH, Shemesh T, Thiagarajan V, Hariadi RF, Anderson KL, Page C, Volkmann N, Hanein D, Sivaramakrishnan S, Kozlov MM, Bershadsky AD. Cellular chirality arising from the self-organization of the actin cytoskeleton. Nat Cell Biol. 2015 Apr;17(4):445-57.
36. Landau, L. D. & Lifshitz, E. M. 1976. Mechanics, 3rd edn. Volume 1. (Course of Theoretical Physics). Oxford: Butterworth- Heinemann.
37. Sato, K., Hiraiwa, T. & Shibata, T. Cell chirality induces collective cell migration in epithelial sheets. *Phys. Rev. Lett.* (2015) 115, 188102.
38. Sato K. Direction-dependent contraction forces on cell boundaries induce collective migration of epithelial cells within their sheet. *Dev Growth Differ.* (2017) Jun;59(5):317-328. doi: 10.1111/dgd.12361
39. Liu YJ, Le Berre M, Lautenschlaeger F, Maiuri P, Callan-Jones A, Heuzé M, Takaki T, Voituriez R, Piel M. Confinement and low adhesion induce fast amoeboid migration of slow mesenchymal cells. *Cell.* (2015) Feb 12;160(4):659-672. doi: 10.1016/j.cell.2015.01.007
40. Sakamoto R, Izri Z, Shimamoto Y, Miyazaki M, Maeda YT. Geometric trade-off between contractile force and viscous drag determines the actomyosin-based motility of a cell-sized droplet. *Proc Natl Acad Sci U S A.* (2022) Jul 26;119(30):e2121147119. doi: 10.1073/pnas.2121147119
41. Camley BA, Zhang Y, Zhao Y, Li B, Ben-Jacob E, Levine H, Rappel WJ. Polarity mechanisms such as contact inhibition of locomotion regulate persistent rotational motion of mammalian cells on micropatterns. Proc Natl Acad Sci U S A. 2014 Oct 14;111(41):14770-5. doi: 10.1073/pnas.1414498111. Epub 2014 Sep 25. PMID: 25258412; PMCID: PMC4205601.
42. Ohta T, Ohkuma T. Deformable self-propelled particles. Phys Rev Lett. 2009 Apr 17;102(15):154101. doi:10.1103/PhysRevLett.102.154101. Epub 2009 Apr 13. PMID: 19518636.
43. Lo CM, Wang HB, Dembo M, Wang YL. Cell movement is guided by the rigidity of the substrate. Biophys J. 2000 Jul;79(1):144-52. doi: 10.1016/S0006-3495(00)76279-5. PMID: 10866943; PMCID: PMC1300921.
44. Taniguchi A, Nishigami Y, Kajiura-Kobayashi H, Takao D, Tamaoki D, Nakagaki

T, Nonaka S, Sonobe S. Light-sheet microscopy reveals dorsoventral asymmetric membrane dynamics of Amoeba proteus during pressure-driven locomotion. Biol Open. 2023 Feb 15;12(2):bio059671. doi: 10.1242/bio.059671. Epub 2023 Feb 23. PMID: 36716104; PMCID: PMC9986612.

45. Moure A, Gomez H. Dual role of the nucleus in cell migration on planar substrates. Biomech Model Mechanobiol. 2020 Oct;19(5):1491-1508. doi: 10.1007/s10237-019-01283-6. Epub 2020 Jan 6. PMID: 31907682.
46. Lecaudey V, Cakan-Akdogan G, Norton WH, Gilmour D. Dynamic Fgf signaling couples morphogenesis and migration in the zebrafish lateral line primordium. *Development.* (2008) Aug;135(16):2695-705. doi: 10.1242/dev.025981
47. Hayakawa M, Hiraiwa T, Wada Y, Kuwayama H, Shibata T. Polar pattern formation induced by contact following locomotion in a multicellular system. *Elife.* (2020) Apr 30;9:e53609. doi: 10.7554/eLife.53609
48. Hirose S, Benabentos R, Ho HI, Kuspa A, Shaulsky G. Self-recognition in social amoebae is mediated by allelic pairs of tiger genes. *Science.* (2011) Jul 22;333(6041):467-70. doi: 10.1126/science.1203903
49. Mlodzik M. Planar polarity in the Drosophila eye: a multifaceted view of signaling specificity and cross-talk. *EMBO J.* (1999) Dec 15;18(24):6873-9. doi: 10.1093/emboj/18.24.6873
50. Mirkovic I, Mlodzik M. Cooperative activities of drosophila DE-cadherin and DN-cadherin regulate the cell motility process of ommatidial rotation. *Development.* (2006) Sep;133(17):3283-93. doi: 10.1242/dev.02468
51. Jenny A. Planar cell polarity signaling in the Drosophila eye. *Curr Top Dev Biol.* (2010); 93:189-227. doi: 10.1016/B978-0-12-385044-7.00007-2

# Figures

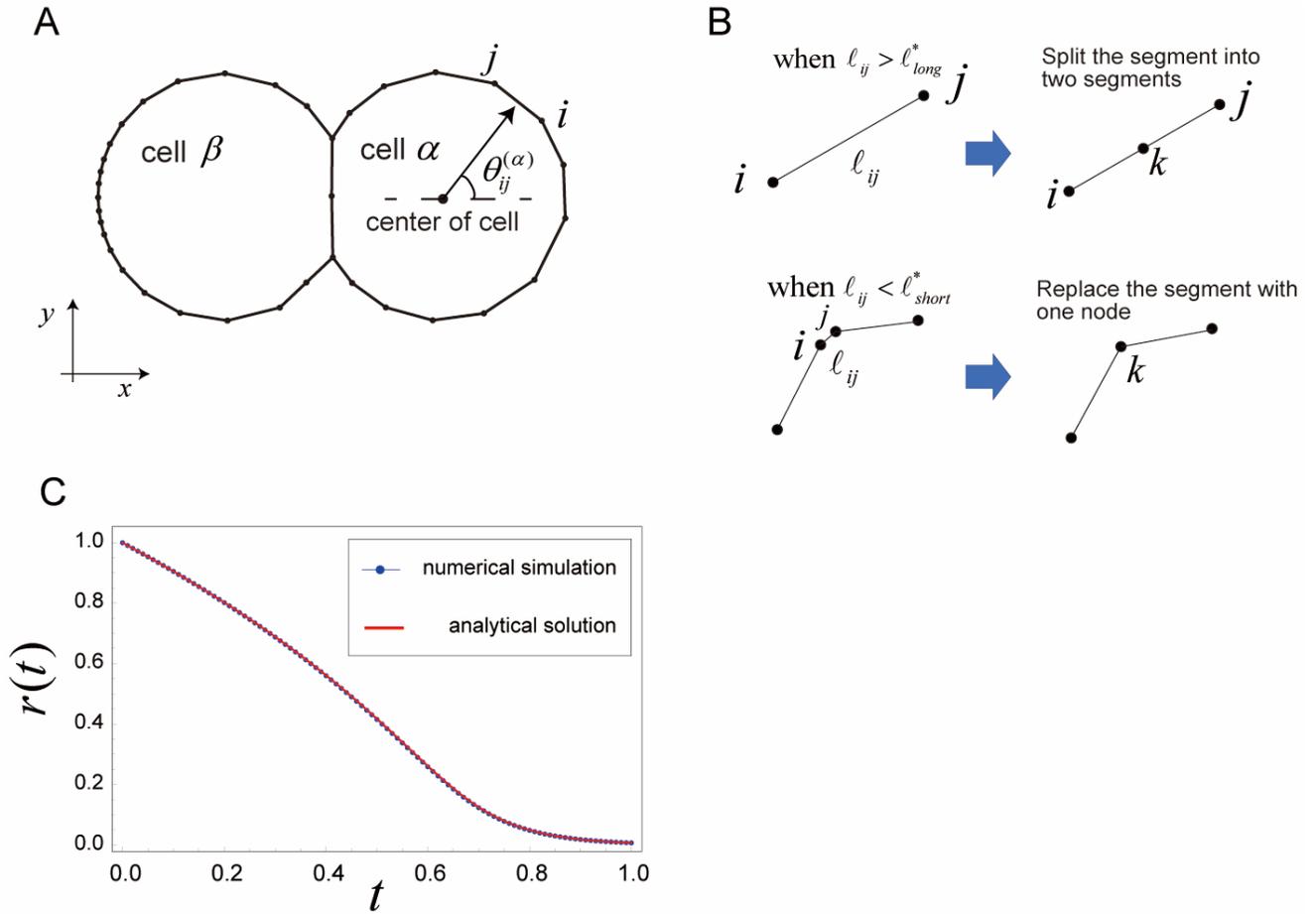

**Figure 1.** Setup for the two-dimensional (2D) cell membrane model of cell migration. (**A**) In this model, cells on a substrate are represented by polygons. If a cell adheres to an adjacent cell, the two cells share the segments and nodes on the joint boundary. The cell surface experiences various mechanical forces that are expressed by $W$ and $U$ in Equations (1) and (2), respectively. These forces are balanced at all times on the cell surface, as shown in Equation (3). (**B**) Replacement rules for cell boundary segments are as follows. When the length $\ell_{ij}$ of segment $ij$ on a cell boundary exceeds a critical length, $\ell^*_{long}$, the segment $ij$ is split into two segments, $ik$ and $kj$, at the next step, creating a new node $k$ at the center of the previous segment $ij$. When $\ell_{ij}$ becomes less than a critical length, $\ell^*_{short}$, the segment $ij$ is replaced with a new node $k$, whose position is the center of the previous segment, $ij$. If appropriate lengths of $\ell^*_{long}$ and $\ell^*_{short}$ are chosen, the cell shape is kept smooth, and movement of the cell surface coincides with that observed under a continuous

limit (*e.g.*, Figure 2B, C.). (**C**) Comparison between the numerical simulation and analytical solution for $r(t)$, where $r(t)$ is the radius of a circular cell that shrinks with constant surface tension $\gamma = 1$, and there are no constraints on area and perimeter of the cell. The parameters used here are as follows: $\eta = 1.0$, $\mu = 0.1$, $K = 0$, $K_p = 0$, $\kappa^{(ijk)} = 0$, and $\Delta t = 1/10000$. The initial configuration for the numerical simulation is a regular 40-sided polygon, with $r = 1$.

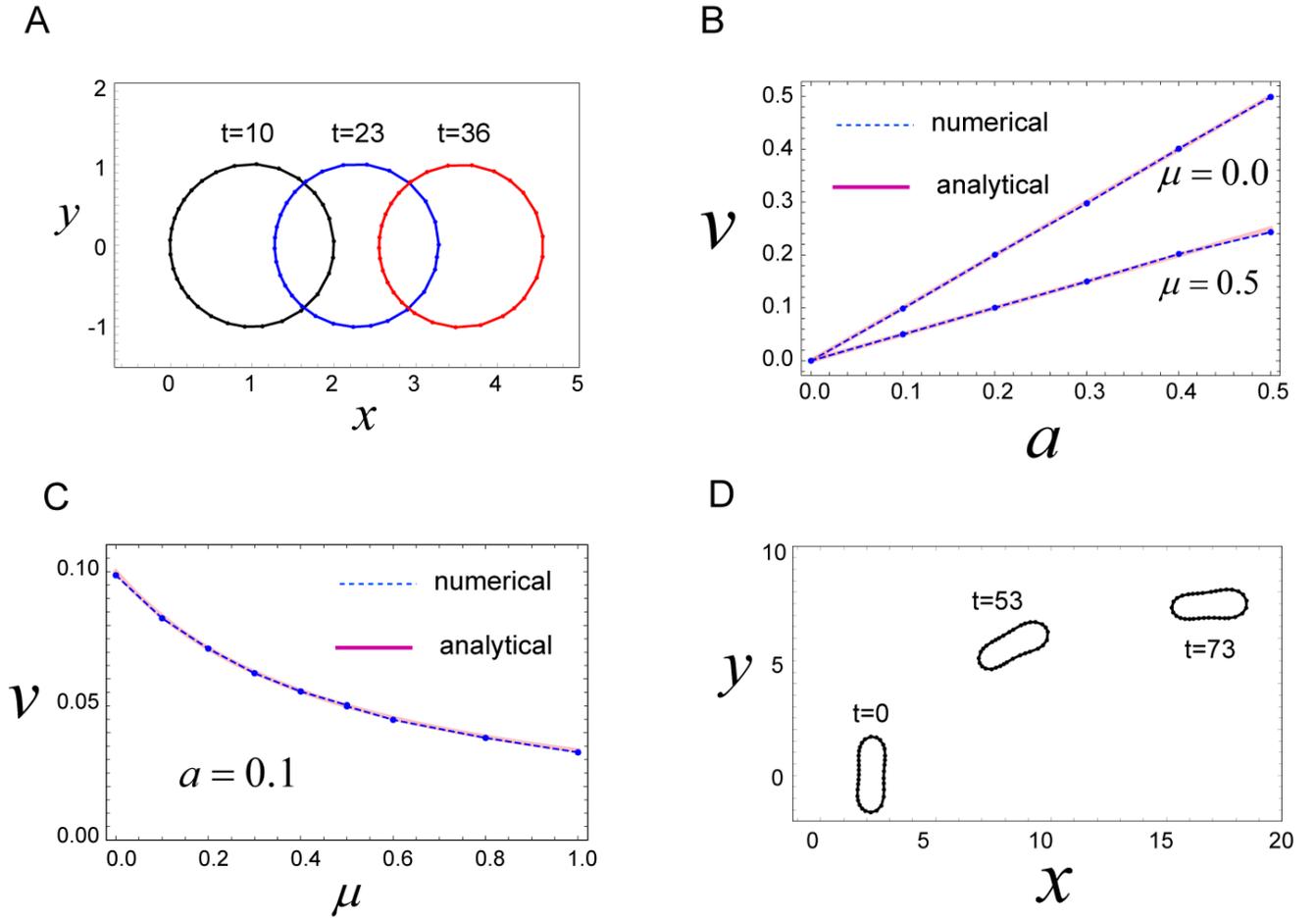

**Figure 2.** Mechanism for migration of a circular and non-circular cell. (**A**) A circular cell migrates in the *x*-direction due to the direction-dependent surface tension in Equation (4), which generates a continuous flow of cell surface from front to rear. The cell movement is maintained as long as the cell has the polarity expressed by Equation (4). Parameters used are as follows: $\mu = 0$, $\gamma_0 = 1.0$, $a = 0.1$, $\ell^*_{long} = 0.31$, $\ell^*_{short} = 0.12$, $K = 100$, $K_p = 0$, and $\kappa^{(ijk)} = 0$. (**B**) The steady speed $v$ of circular cell movement as a function of $a$ that represents the degree of polarization of the cell (see Equation (4)). The numerical results

(dashed lines) and analytical results (red lines) show good agreement with one another, indicating that our discrete model, described by Equations (1)–(3), coincides with a continuous model if we choose appropriate values for $\ell^*_{long}$ and $\ell^*_{short}$. Parameters used are as follows: $\gamma_0 = 1.0$, $\ell^*_{long} = 0.31$, $\ell^*_{short} = 0.12$, $K = 100$, $K_p = 0$, and $\kappa^{(ijk)} = 0$. (**C**) The steady speed $v$ of circular cell movement as a function of $\mu$, which is the friction coefficient for internal friction of the cell surface, defined in Equation (1). The analytical results (red curve) and numerical results (dashed curve) are in agreement with one another. Parameters used are as follows: $\gamma_0 = 1.0$, $a = 0.1$, $\ell^*_{long} = 0.31$, $\ell^*_{short} = 0.12$, $K = 100$, $K_p = 0$, and $\kappa^{(ijk)} = 0$. (**D**) A non-circular cell also migrates due to the direction-dependent surface tension in Equation (4). In the final state, the long axis of the cell is aligned in the $x$-axis. Parameters used here are $\mu = 0$, $\gamma_0 = 0$, $a = 0.2$, $\ell^*_{long} = 0.4$, $\ell^*_{short} = 0.15$, $K = 1.0$, $K_p = 1.0$, $\kappa^{(ijk)} = 0.1$, and $L_0 = 2.5\pi$.

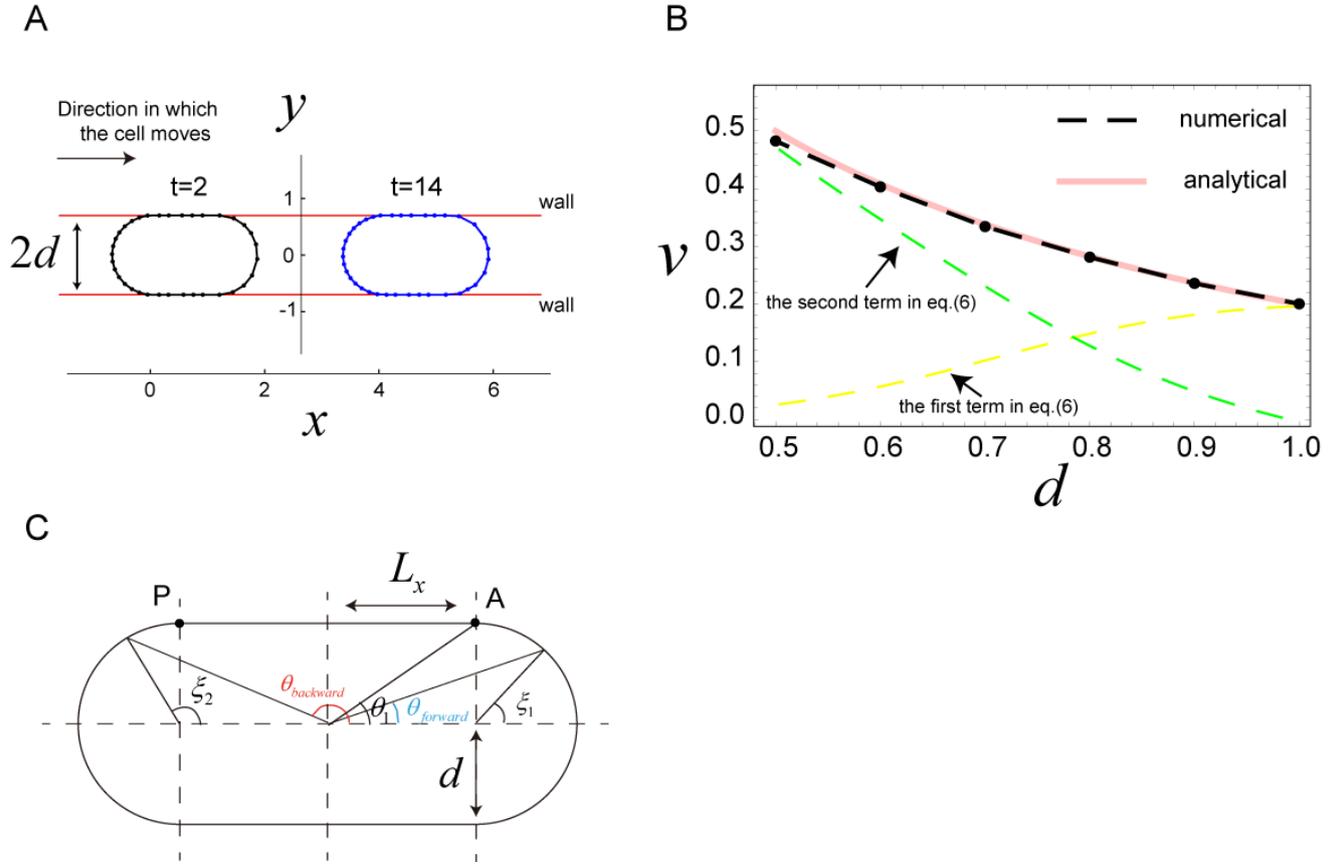

**Figure 3.** Migration of a cell sandwiched by two walls. (**A**) Snapshots of a cell sandwiched by two walls, which migrates in the *x*-direction due to surface flow resulting from the direction-dependent surface tension in Equation (4). The distance between the two walls is $2d$; a decrease in $d$ increases migration speed, $v$, of the cell (see panel B). Parameters used are as follows: $d = 0.7$, $\mu = 0$, $\gamma_0 = 1.0$, $a = 0.2$, $\ell^*_{long} = 0.31$, $\ell^*_{short} = 0.12$, $K = 100$, $K_p = 0$, and $\kappa^{(ijk)} = 0$. (**B**) The steady speed, $v$, of a cell sandwiched by walls as a function of $d$. The dashed black curve shows the numerical results, and the red curve represents the analytical results from Equation (6), which are in good agreement. The yellow and green dashed curves are the first and second terms in Equation (6), respectively. Parameters used here are the same as in panel A, except that $d$ is varied. (**C**) Geometrical explanation of the quantities $\theta_1$, $\theta_{forward}$, $\theta_{backward}$, $\xi_1$, and $\xi_2$ in Equation (6).

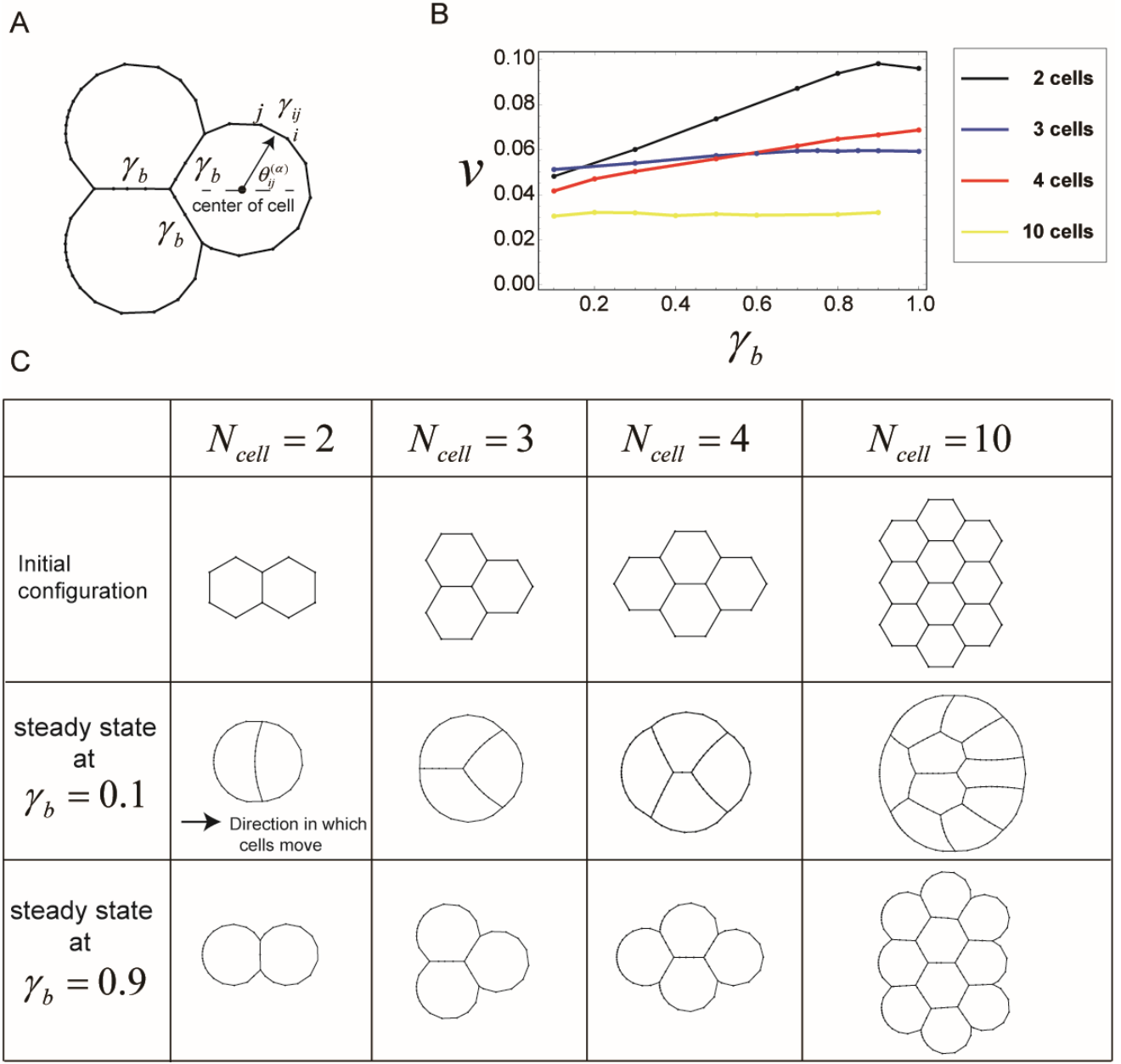

Figure 4. Cells in a cluster also migrate due to direction-dependent surface tension. (A) Setup for the surface tension of cells in a cluster. The boundaries between any two cells in the cluster have a constant surface tension $\gamma_b$, whereas the other cell surfaces, which are outside of the cluster, have the direction-dependent surface tension expressed by Equation (4). (B) The steady speed $v$ of cell cluster migration as a function of $\gamma_b$, for each cell number $N_{cell}$. When $\gamma_b$ is increased, cluster migration speed also tends to increase. Parameters used here

are as follows: $\mu = 0$, $\gamma_0 = 1.0$, $a = 0.1$, $\ell^*_{long} = 0.6$, $\ell^*_{short} = 0.08$, $K = 100$, $K_p = 0$, and $\kappa^{(ijk)} = 0$. (C) Shapes of cells in the cluster during steady state migration for each $N_{cell}$ and $\gamma_b$.

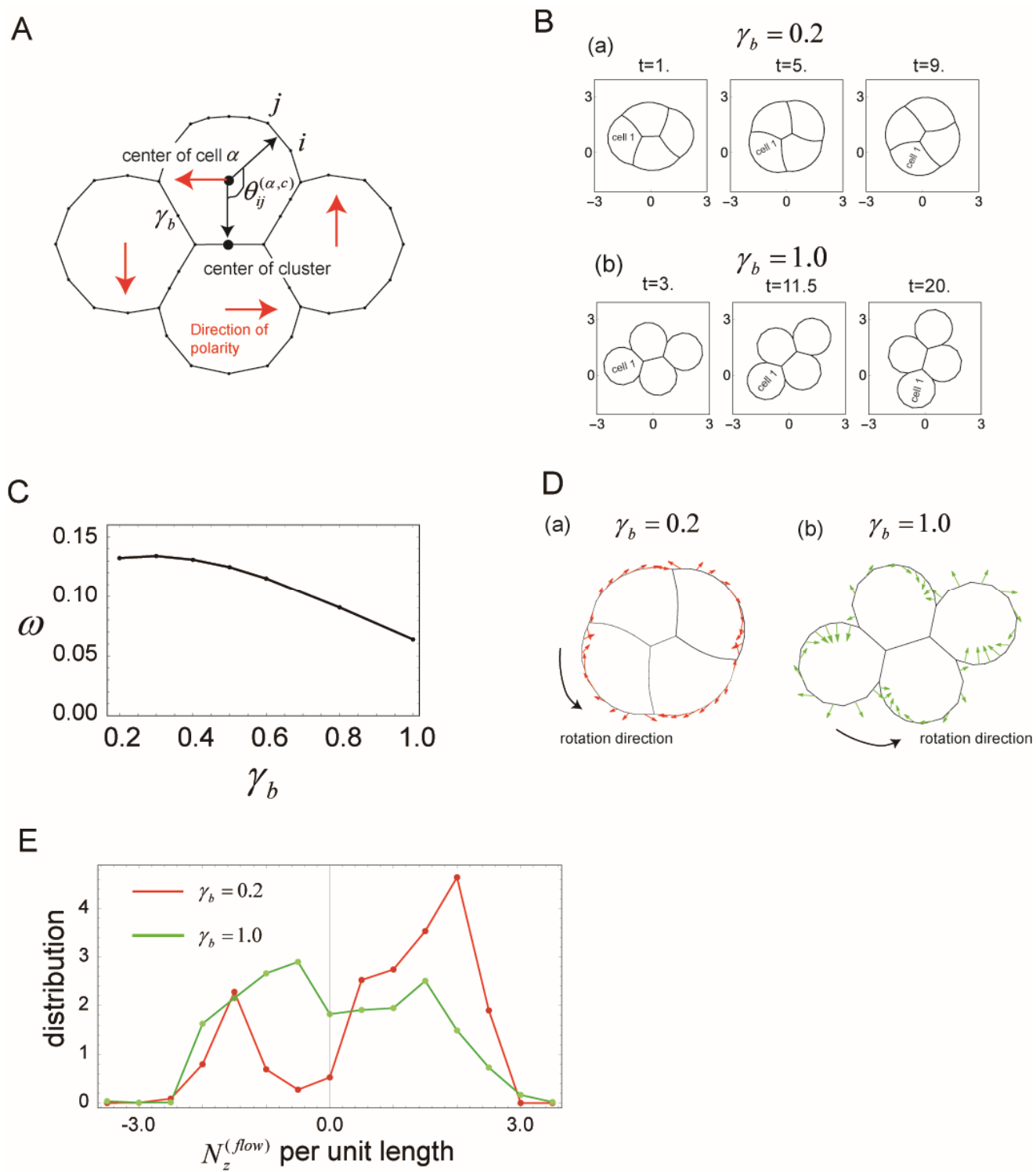

Figure 5. Cells in a cluster rotate when the polarity of each cell is tilted with respect to the cluster center. (A) Geometrical explanation of the angle $\theta_{ij}^{(a,c)}$ used in Equation (13). Polarity of each cell in the cluster is indicated by the red arrow, which has a direction

perpendicular to the vector from the cell center to the center of the cluster. (**B**) Snapshots of the cell cluster, rotating counterclockwise about the center of the cluster due to the direction-dependent surface tension in Equation (13). Parameters used are as follows: $\mu = 0$, $\gamma_0 = 1.0$, $a = 0.1$, $\ell^*_{long} = 0.6$, $\ell^*_{short} = 0.08$, $K = 100$, $K_p = 0$, and $\kappa^{(ijk)} = 0$. In (a), $\gamma_b = 0.2$ and in (b), $\gamma_b = 1.0$. (**C**) The angular velocity, $\omega$, of cluster rotation as a function of $\gamma_b$. (**D**) Velocities of the cell surface during rotation. (a) Red arrows indicate the cell surface velocities for $\gamma_b = 0.2$. (b) Green arrows indicate the cell surface velocities for $\gamma_b = 1.0$. The velocities for $\gamma_b = 0.2$ are oriented closer to the tangent of the cell surface, compared with those for $\gamma_b = 1.0$. (**E**) Distribution of the $z$-component of the torque generated by surface flow, which is induced by the direction-dependent surface tension in Equation (13). When $\gamma_b = 0.2$, stronger torques that rotate the cluster more rapidly are generated, compared with the torques produced when $\gamma_b = 1.0$.

# 8 Supplementary Material

We provide fourteen movies S1-S14 and one text that includes Appendices A-D and supplementary figures S1 and S2 as the supplementary materials. You can download these files on the website of Frontiers.